\documentclass[letter]{aa} 
\usepackage{graphicx}
\usepackage{txfonts}
\def\vsin{\hbox{$v \sin i$}}  
  
\def\kms{\hbox{km\,s$^{-1}$}}  
\def\ms{\hbox{m\,s$^{-1}$}}

\def\em{\it}  
  
\def\degr{\hbox{$^\circ$}}  
\def\rpd{\hbox{rad\,d$^{-1}$}}   
   
\def\omeq{\hbox{$\Omega_{\rm eq}$}}   
\def\dom{\hbox{$d\Omega$}}   
\def\kis{\hbox{$\chi^2$}}   
\def\kisr{\hbox{$\chi^2_{\rm r}$}}   
\def\drot{\hbox{differential rotation}}

\def\hdc{\hbox{HD~190771}}   
\def\msun{\hbox{$M_\odot$}}

\begin{document}
   \title{A polarity reversal in the large-scale magnetic field of the rapidly rotating sun HD~190771}
   \titlerunning{Magnetic polarity reversal of HD~190771}

   \author{P. Petit
          \inst{1}
          \and
          B. Dintrans
          \inst{1}  
          \and
          A. Morgenthaler
          \inst{1}
          \and
          V. Van Grootel
          \inst{1}
          \and
          J. Morin
          \inst{1}
          \and
          J. Lanoux
          \inst{2}
          \and  
          M. Auri\`ere
          \inst{1} 
          \and
          R. Konstantinova-Antova
          \inst{3}
          }

   \offprints{P. Petit}

   \institute{
   Laboratoire d'Astrophysique de Toulouse-Tarbes, Universit\'e de Toulouse, CNRS, France \\ 
   \email{petit@ast.obs-mip.fr, dintrans@ast.obs-mip.fr, auriere@ast.obs-mip.fr,\\ amorgent@ast.obs-mip.fr, jmorin@ast.obs-mip.fr, vvangroo@ast.obs-mip.fr}
               \and
             Centre d'Etude Spatiale des Rayonnements, Universit\'e de Toulouse, CNRS, France \\ 
             \email{joseph.lanoux@cesr.fr}
            \and
             Institute of Astronomy, Bulgarian Academy of Sciences, 72 Tsarigradsko shose, 1784 Sofia, Bulgaria\\
             \email{antovi@astro.bas.bg}
             }

   \date{Received ??; accepted ??}

 
  \abstract
   {}
   {We investigate the long-term evolution of the large-scale photospheric magnetic field geometry of the solar-type star HD~190771. With fundamental parameters very close to those of the Sun except for a shorter rotation period of 8.8~d, HD~190771 provides us with a first insight into the specific impact of the rotation rate in the dynamo generation of magnetic fields in 1~\msun\ stars.}
   {We use circularly polarized, high-resolution spectra obtained with the NARVAL spectropolarimeter (Observatoire du Pic du Midi, France) and compute cross-correlation line profiles with high signal-to-noise ratio to detect polarized Zeeman signatures. From three phase-resolved data sets  collected during the summers of 2007, 2008, and 2009, we model the large-scale photospheric magnetic field of the star by means of Zeeman-Doppler imaging and follow its temporal evolution.}
   {The comparison of the magnetic maps shows that a polarity reversal of the axisymmetric component of the large-scale magnetic field occurred between 2007 and 2008, this evolution being observed in both the poloidal and toroidal magnetic components. Between 2008 and 2009, another type of global evolution occured, characterized by a sharp decrease of the fraction of magnetic energy stored in the toroidal component. These changes were not accompanied by significant evolution in the total photospheric magnetic energy. Using our spectra to perform radial velocity measurements, we also detect a very low-mass stellar companion to \hdc.}
   {
   }

   \keywords{stars: individual: HD 190771 -- stars: magnetic fields -- stars: late-type -- stars: rotation -- stars: atmospheres -- stars: activity}

   \maketitle

\section{Introduction}

All rotating Sun-like stars exhibit spectral features indicative of magnetic fields in their atmospheres. Chromospheric emission (as measured for instance in the cores of CaII H \& K spectral lines) is often taken to be a good magnetic tracer, assuming that the correlation observed locally in the Sun between photospheric magnetic field and chromospheric flux (Schrijver et al. 1989) holds for other stars as well. For a few tens of stars, time-series covering several decades are now available and provide information about the existence and length of magnetic cycles that different types of cool stars generate (Baliunas et al. 1995, Hall et al. 2007a). From this long-term monitoring, several trends can be inferred. First, the chromospheric flux increases with the rotation rate, which is indicative of a more efficient field generation by dynamo action whenever fast rotation is present. Another observation is that all active stars do not undergo smooth activity cycles, as the Sun does most of the time. Among the most active dwarfs (which are also younger and rotate much more rapidly than the Sun), stellar activity has a tendency to fluctuate erratically, while regular cycles are instead observed in slowly rotating, older stars such as the Sun.  It also appears as if the long-term magnetic variability of Sun-like stars is very sensitive to fundamental stellar parameters, so that stars very similar to the Sun can have a different magnetic behaviour. While often being considered an excellent solar analogue, 18 Sco has been reported to follow an activity cycle shorter than solar, with a period of $\approx$7 years (Hall et al. 2007b).

Since many stars experience a series of activity maxima and minima in a similar way to the Sun, a natural step is to determine whether these oscillations in chromospheric flux are associated with global polarity reversals of the large-scale magnetic field, as observed on the Sun between two successive solar minima. For the Sun, the global component of the magnetic field displays a strength of a few Gauss (e.g., Sanderson et al. 2003), is dominated by a dipole that is almost aligned with the spin axis when the Sun is close to the minimum of its activity cycle, and is organized in a more complex multipolar geometry around solar maximum. The global component of the magnetic field has been observed on other cool active stars (Petit et al. 2005, Petit et al. 2008, hereafter P08). The observation of a preliminary sample of 1 M$_\odot$ stars by P08 suggests that rapid rotation has the effect of generating a large-scale toroidal field at the stellar surface, the majority of the surface magnetic energy being concentrated in the toroidal field whenever the rotation period is shorter than about 2 weeks, as also found by 3-D MHD simulations of stellar dynamos (Brown et al. 2009). Our first study did not, however, provide us with any information about the variety of configurations that these non-solar magnetic geometries can assume with time.  

In the present paper, we concentrate on \hdc, the most rapid rotator in the stellar sample first presented by P08, to investigate the temporal evolution of its large-scale magnetic field. We first describe the new data sets obtained in 2008 and 2009, as well as the procedure used to reconstruct magnetic maps from the time-series of high signal-to-noise line-profiles. We then compare the maps obtained in 2007 (see P08), 2008, and 2009 and discuss the information we obtain about the nature of magnetic cycles in rapidly rotating solar analogues.

\section{Observations and data modelling}
\label{sect:data}

\subsection{Instrumental setup, data reduction, and multi-line extraction of Zeeman signatures}

\onltab{1}{
\begin{table}
\caption[]{Journal of observations for 2008 and 2009 (2007 observations are described in P08).}
\begin{tabular}{cccccc}
\hline
Year & Julian Date & exp. time & $\sigma_{\rm LSD}$ & rot. phase \\
 & (2,450,000+) & sec. & $10^{-5}I_{c}$ & & \\
\hline
2008 & 4696.49 & 1600.0 & 6.7268   & 0.6133 \\
 &4701.41 & 1600.0 & 9.2876   & 0.1726 \\
 &4702.48 & 1600.0 & 3.8056   & 0.2936 \\
 &4703.45 & 1600.0 & 3.8787   & 0.4040 \\
 &4704.50 & 1600.0 & 4.0409   & 0.5237 \\
 &4725.39 & 960.0   & 5.0916   & 0.8975 \\
 &4725.41 & 960.0   & 5.1507   & 0.8991 \\
 &4725.42 & 960.0   & 5.1279   & 0.9006 \\
 &4725.43 & 960.0   & 5.1300   & 0.9021 \\
 &4726.38 & 1600.0 & 4.3341   & 0.0093 \\
 \hline
 2009 &4984.61 & 1600.0 & 4.9114   & 0.3538 \\
 &4985.61 & 1600.0 & 4.8112   & 0.4673 \\
 &4986.57 & 1600.0 & 4.8541   & 0.5763 \\
 &4995.55 & 1600.0 & 11.8554 & 0.5970 \\
 &5002.59 & 1600.0 & 4.4350   & 0.3965 \\
 &5002.61 & 1600.0 & 9.0171   & 0.3988 \\
 &5005.58 & 2400.0 & 4.4490   & 0.7373 \\
 &5006.58 & 2400.0 & 14.8898 & 0.8504 \\
 &5010.54 & 2400.0 & 4.5288   & 0.3001 \\
 &5011.53 & 2400.0 & 4.8640   & 0.4129 \\
 &5013.57 & 1600.0 & 4.9603   & 0.6446 \\
 &5017.61 & 2400.0 & 5.1851   & 0.1043 \\
 &5018.62 & 2400.0 & 4.8179   & 0.2188 \\

\hline
\end{tabular}
\noindent Notes: From left to right, we list the year of observation, the Julian date, the exposure time, the error-bar in Stokes V LSD profiles, and the phase of the rotational cycle at which the observation was made, taking the same rotation period and phase origin as P08.
\label{tab:obs}
\end{table}
}

The new data sets were obtained with the 2-m Telescope Bernard Lyot (Observatoire du Pic du Midi, France), using the NARVAL spectropolarimeter. The instrumental setup and data reduction procedure are identical to those presented by P08. Observations consist of high-resolution spectra (R = 65,000) covering the entire optical domain (from 370 to 1,000 nm) and providing simultaneous recording of the stellar light as unpolarized (Stokes I) and circularly polarized (Stokes V) spectra.  In 2008, a total of 10 spectra were recorded between August 15 and September 16 (Table \ref{tab:obs}). Another set of 13 spectra were acquired in 2009, between June 02 and July 05. The data were processed with LibreEsprit, the automatic reduction software developed for NARVAL (Donati et al. 1997).  

For each spectrum, both Stokes I \& V parameters are processed using the LSD technique (Least-Squares Deconvolution, Donati et al. 1997). Using a line-mask defined by a solar photospheric model, we compute LSD line profiles using about 5,000 spectral lines. The multiplex gain in signal-to-noise ratio between the raw spectra and the LSD mean profiles is about 30, reducing the noise to between 4 and $15\times 10^{-5}I_{c}$, where $I_{c}$ represents the intensity of the continuum. In these new data sets, the noise is higher and also more variable than observed in our 2007 data, owing to poor transparency that significantly affected part of the 2008 and 2009 observing runs.

\begin{figure}
\centering
\mbox{
\includegraphics[height=8cm]{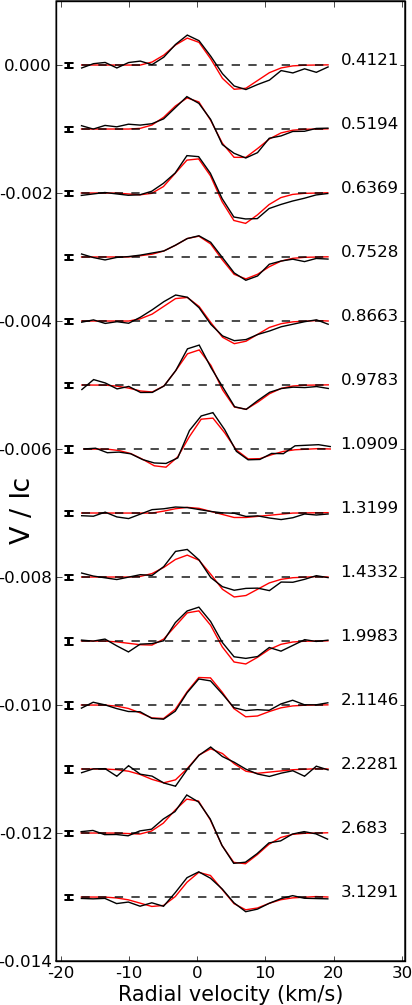}
\includegraphics[height=8cm]{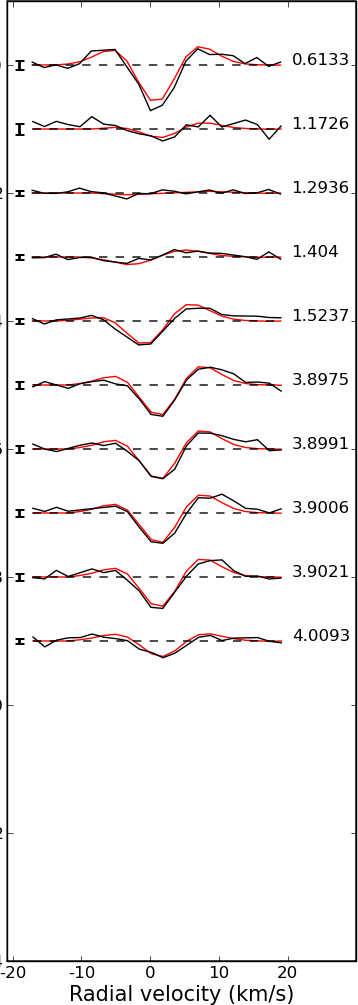}
\includegraphics[height=8cm]{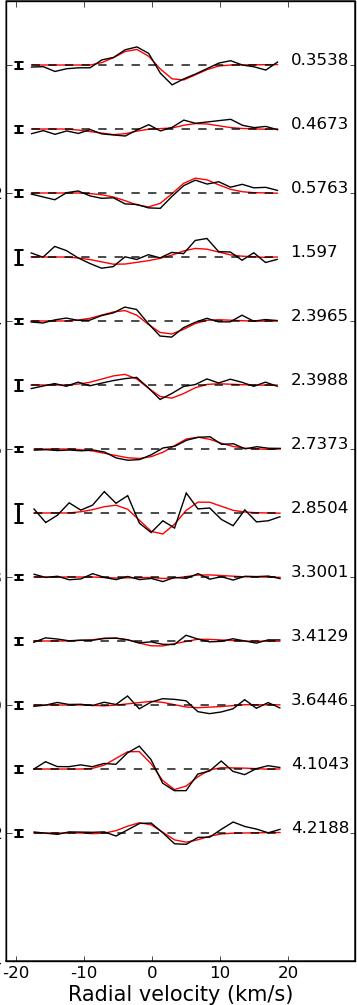}}
\caption{Normalized Stokes V profiles of HD 190771 for 2007, 2008 and 2009 (from left to right), after correction for the mean radial velocity of the star. Black lines represent the data and red lines correspond to synthetic profiles of our magnetic model. Successive profiles are shifted vertically for display clarity. Rotational phases of observations are indicated in the right part of the plot and error bars are illustrated on the left of each profile.}
\label{fig:stokesv}
\end{figure}

The sequences of Stokes V LSD line profiles are plotted in Fig. \ref{fig:stokesv}, after removal of a mean radial velocity of -26.7 and -26.5 \kms\ in 2008 and 2009, respectively. Zeeman signatures are evident in a majority of the observations. In 2008, they display several similarities with the signatures recorded in the 2007 data set. We first note the maximum recorded amplitude of the Stokes V signal of  $7\times 10^{-4} I_c$ (of about 10 times the noise level), against $6\times 10^{-4} I_c$ one year earlier. We also note that the shape of the Zeeman signatures evolves with time, the signatures being sometimes mostly symmetric about the line center (rotational phase 0.6133 in 2008, 0.0909 in 2007), sometimes almost anti-symmetric (phase 0.5237 in 2008, 0.4121 in 2007), and sometimes almost flat (phases 0.2936 in 2008, 0.3199 in 2007). All of these recognizable line-patterns are evident in 2008 with a sign reversal in the Zeeman signatures compared to 2007 observations. No similar sign change is observed in the Stokes V profiles of other cool active stars monitored during the same two runs, ensuring that it is very unlikely that a possible sign switch has been introduced by mistake during data reduction, affecting all data collected during one of the runs. In 2009, another type of evolution was observed relative to 2008, symmetric Stokes V signatures being absent in the latest data set, and antisymmetric signatures of both polarities being observed all the time, except a possibly symmetric (although noisy) signature at phase 0.8504.

\subsection{Magnetic mapping}

\begin{figure*}
\centering
\mbox{\includegraphics[height=6.5cm]{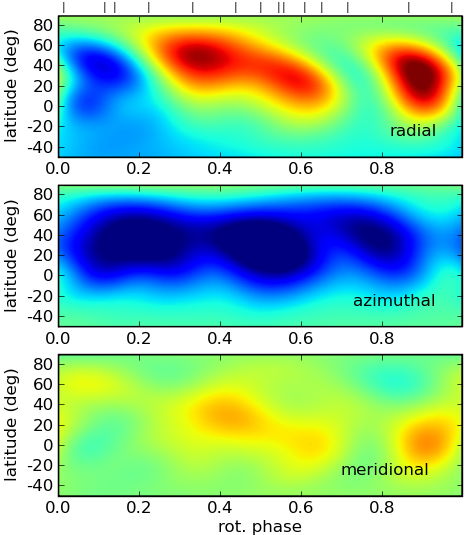}
\includegraphics[height=6.5cm]{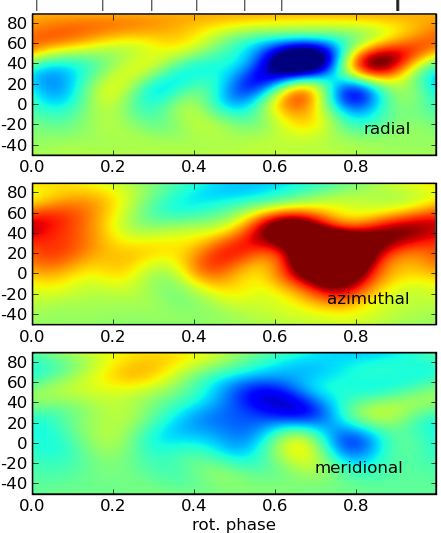}
\includegraphics[height=6.5cm]{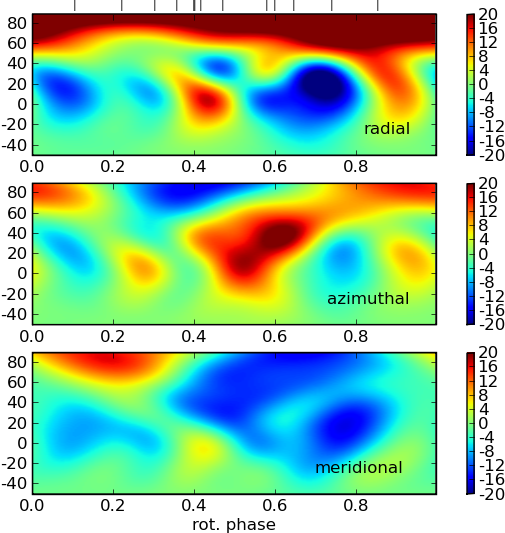}}
\caption{Magnetic maps of HD 190771, derived from 2007, 2008, and 2009 observations in the left, middle, and right columns, respectively. The map on the left is plotted after P08. For each data set, the 3 charts illustrate the field projection onto one axis of the spherical coordinate frame with, from top to bottom, the radial, azimuthal, and meridional field components. The magnetic field strength is expressed in gauss and the rotational phases of observation are indicated as vertical ticks above each epoch.}
\label{fig:map}
\end{figure*}

The sets of profiles are used to model the large-scale magnetic geometry of the star, using the tomographic inversion technique of Zeeman-Doppler Imaging (Donati \& Brown 1997, Donati et al. 2006). To compute synthetic Stokes V line profiles, we develop a model of a stellar surface divided into a grid of pixels, each pixel being associated with a local Stokes I and V profile. For a given magnetic field strength and orientation of each pixel, local Stokes V profiles are calculated based on the weak field assumption, i.e., assuming that Stokes V  is proportional to $g.\lambda_0^2 .B_\parallel .\partial I / \partial \lambda$, where $\lambda_0$ is the average wavelength of the LSD profile (about 534 nm), $B_\parallel$ is the line-of-sight projection of the local magnetic field vector, and $g$ is the effective Lande factor of the LSD profile, which is equal to 1.2. We further assume that there is no large-scale temperature inhomogeneities over the stellar surface, so that Stokes I profiles are locally the same over the whole photosphere. For each pixel, we impose a Gaussian shape on the local synthetic Stokes I profile, which has a FWHM equal to 10.1 \kms\ (which provides the best fit to the observed series of Stokes I profiles). At a given rotational phase, the amplitude of the local Stokes profiles located on the visible hemisphere of the star are then weighted according to a linear limb-darkening coefficient equal to 0.75 (see P08) and their wavelength location is shifted according to the line-of-sight velocity of the pixel, assuming that \vsin=4.3 \kms\ (Valenti \& Fischer 2005) and $i=50$\degr\ (P08).

Synthetic Stokes V profiles are computed for all observed rotation phases and compared to the data sets. The data adjustment is iterative and based on a maximum entropy algorithm (Skilling \& Bryan 1984). The surface magnetic field is projected onto a spherical harmonics frame (Donati et al. 2006), where the magnetic field geometry is divided between a poloidal and toroidal component (Chandrasekhar 1961). As in P08, we limit the spherical harmonics expansion to $\ell \leq 10$, after checking that increasing even more $\ell$ does not provide a superior data adjustment. We finally assume that the star is not rotating as a rigid body, but experiences a latitudinal shear that we simply model as $\Omega(\theta)=\Omega_{\rm eq} - d\Omega .\sin^2(\theta)$, where $\theta$ is the stellar latitude, $\Omega_{\rm eq}$ is the rotational rate of the equator and $d\Omega$ is the difference in rotation rate between polar and equatorial regions. We use the new data sets to obtain estimates of the \drot\ parameters (Table \ref{tab:dynamo}), following the method described by Petit et al. (2002). For 2008 and 2009, we obtain a shear level of $d\Omega=0.12$~\rpd, in very good agreement with the value obtained by P08. Values derived for \omeq\ are also in excellent agreement in 2007 and 2008 (with \omeq$=0.71\pm 0.01$~\rpd), but the measurement for 2009 provides us with a different estimate (\omeq$=0.66\pm 0.01$~\rpd). This apparent discrepancy may simply reflect the uncertainties in this parameter. Errorbars listed in Table \ref{tab:dynamo} are directly derived from the \kis\ map in the \omeq-\dom\ plane and might be underestimated, as suggested by Petit et al. (2002). 

Using this procedure, the spectropolarimetric data are adjusted at a reduced \kis\ equal to 1.1 and 0.9, in 2008 and 2009, respectively (Fig. \ref{fig:stokesv}). The \kisr\ value for 2008 equals that obtained one year earlier, and the slightly smaller \kisr\ achieved in 2009 is partly due to the higher relative noise in the data. The reconstructed magnetic topologies are illustrated in Fig. \ref{fig:map}, together with the magnetic map obtained in 2007 (P08). In Table \ref{tab:dynamo}, we list several numerical quantities derived from the spherical harmonics coefficients defining the magnetic geometries. As in P08, errorbars listed in Table \ref{tab:dynamo} are estimated by reconstructing a set of magnetic maps using different input parameters for the inversion code (with individual parameters being varied over the width of their own errorbars). We note that the largest variations in the output quantities are generally obtained by varying the stellar inclination, because of the relatively large uncertainty in this specific parameter. 

\subsection{Temporal evolution in the large-scale magnetic field}

\begin{table*}
\caption[]{Magnetic quantities derived from the set of magnetic maps. }
\begin{center}
\begin{tabular}{ccccccccccc}
\hline
fractional year               & $v_{\rm r}$          & $B_{\rm mean}$ & pol. en.        & dipole            & quad.             & oct.                  & axi.                   & \omeq                    & \dom                    & $\log R'_{\rm HK}$ \\
                        & (\kms)                          & (G)                        & (\% tot)         & (\% pol)         & (\% pol)          & (\% pol)          &(\% tot)              & (\rpd)                        &  (\rpd)                  &                                   \\
\hline
2007.59               & $-26.86 \pm 0.03$ & $51\pm 6$          & $34\pm 1$   & $43\pm 8$   & $20\pm 2$     & $23 \pm 4$    & $73\pm 3$     & $0.71 \pm 0.01$ & $0.12\pm 0.03$  & -4.47 \\
2008. 67               & $-26.72 \pm 0.04$ & $59\pm 3$          & $39\pm 3$   & $36\pm 8$   & $18\pm 2$     & $19 \pm 4$    & $61 \pm 3$     & $0.71 \pm 0.01$ & $0.12 \pm 0.03$ & -4.47 \\
2009. 47               & $-26.48 \pm 0.03$ & $58\pm 8$          & $81\pm 2$   & $23\pm 7$   & $40\pm 2$     & $21 \pm 2$    & $36 \pm 12$  & $0.66 \pm 0.01$ & $0.12 \pm 0.02$ & -4.48 \\
\hline
\end{tabular}
\end{center}
\noindent Notes: We list the stellar radial velocity (with its associated $rms$), the mean unsigned magnetic field ($B_{\rm mean}$), the fraction of the large-scale magnetic energy reconstructed in the poloidal field component, the fraction of the {\em poloidal} magnetic energy in the dipolar ($\ell = 1$), quadrupolar ($\ell = 2$), and octopolar ($\ell = 3$) components, and the fraction of energy stored in the axisymmetric component ($m=0$). We also list the differential rotation parameters \omeq\ and \dom. The last column contains the $\log R'_{\rm HK}$ values derived from our sets of Stokes I spectra. Values for 2007 are taken from P08, except for the $\log R'_{\rm HK}$ value which was recalculated using a new calibration of NARVAL measurements against Mount Wilson estimates (Wright et al. 2004), involving 29 solar-type dwarfs.
\label{tab:dynamo}
\end{table*}

Between 2007 and 2008, the most striking evolution in the magnetic field distribution is a polarity reversal of the large-scale field.  In the ZDI maps of Fig. \ref{fig:map}, this change is mostly visible in the azimuthal component of the magnetic vector. A more quantitative way of estimating the details of this sign switch consists of tracking its origin in the evolution of the complex spherical harmonics coefficients $\alpha_{\ell,m}$,  $\beta_{\ell,m}$, and  $\gamma_{\ell,m}$ (defined by Donati et al., 2006). Because of the uncertainty in the stellar rotation period that prevents us from comparing, at 1-year intervals, magnetic features that manifest themselves at specific rotation phases, we choose to limit our comparison to axisymmetric features (defined by modes with $m=0$). We observe that all coefficients $\alpha_{\ell,0}$,  $\beta_{\ell,0}$, or  $\gamma_{\ell,0}$ with a magnetic amplitude greater than 1 gauss (which only concerns modes with $\ell \le 4$) have a different sign in both years, with the marginal exception of $\gamma_{3,0}$. Another noticeable temporal evolution concerns the fraction of magnetic energy stored in the axisymmetric field component (all spherical harmonics modes with $m=0$), 61\% of the large-scale magnetic energy being in axisymmetric modes in 2008, compared to 73\% in 2007.  Another difference is the fraction of magnetic energy contained in modes with $\ell > 3$ (14\% and 27\%, for 2007 and 2008, respectively, if the poloidal component is considered alone; 5\% and 13\% if the toroidal field is also taken into account), suggesting that the field distribution is more complex in 2008. 

A similar comparison between the 2008 and 2009 maps also reveals a striking evolution in the field geometry. The main change in the field distribution shows up as a much higher fraction of the magnetic energy being stored in the poloidal field component (about 80\% in 2009, against less than 40\% in the two other data sets). This evolution occurs together with an increased level of non-axisymmetry in the field distribution, with only 36\% of the magnetic energy showing up in modes with $m=0$. We also note that the poloidal component of the field is dominated by the quadrupolar terms in 2009, while the dipole was predominant in 2007 and 2008. The polarity of the global field is the same in 2008 and 2009 (the only spherical harmonics coefficients with both $m=0$ and an amplitude greater than 1 gauss having switched sign in the meantime are $\alpha_{1,0}$ and $\gamma_{3,0}$). 

\section{Discussion}
\label{sect:discussion}

From new observations of \hdc\ completed in 2008 and 2009, we have obtained information about the magnetic properties of rapidly rotating 1 \msun\ stars, and provided fresh insight into the temporal changes affecting their magnetic geometries. As already outlined by P08,  \hdc\ exhibits a large-scale toroidal component in its photospheric magnetic field, an element that is absent on the solar surface but whose incidence appears to be common among solar analogues with sufficiently high rotation rates. Analyses of the new data sets show that the toroidal field can experience dramatic changes over a period of a few months. At least two types of evolution are evident, with the possibility of polarity switches (between 2007 and 2008), as well as the possible conversion of magnetic energy from the toroidal to the poloidal field component (from 2008 to 2009). These changes occur with a total magnetic energy that remains approximately constant, as can be seen from the mean (unsigned) magnetic field or from the stellar chromospheric emission (Table \ref{tab:dynamo}).

Similarly rapid evolution of a photospheric magnetic field was reported for $\tau$~Bootis (Fares et al. 2009), where a close-by planet orbiting the star was associated with a significant tidal coupling between the star and the planet. While no low-mass companion has been reported for \hdc, radial velocity ($v_r$) measurements listed in Table \ref{tab:dynamo} show that \hdc\ is not a single star, since the observed variations in $v_r$, of the order of 400 \ms, are much larger than the radial velocity accuracy of NARVAL (about 30 \ms, Moutou et al. 2007). Older radial velocity estimates by Nidever et al. (2005), dating back to 2000, confirm the existence of a long-term trend, with a mean radial velocity of 25.063 \kms\ at that time. Despite the poorly sampled time-series at our disposal, a preliminary estimate of the companion mass can be performed, assuming for simplicity a single companion, a circular orbit, and an orbital plane orthogonal to the stellar rotation axis. By doing so, the best fit is obtained for an orbital period of  14.2 yr, an orbital radius of 5.9 AU, and a companion mass of 0.1 \msun.  We note that a short orbital period (of a few days) can be ruled out by our data, since no significant variations in  $v_r$ are observed within the timespan of our individual data sets, apart from weaker activity-induced fluctuations. Although rough, this first estimate suggests that tidal effects generated by the presence of the companion remain weak (due to the absence of synchronized orbital and rotation periods) and are unlikely to affect significantly the dynamics of the outer convective layers of \hdc. In this context, the fast magnetic field evolution of the star is probably controlled by other fundamental parameters, the rapid rotation being the most obvious deviation from a strict solar situation. In turn, and although $\tau$~Bootis also differs from the Sun in terms of its higher mass ($1.33\pm0.11$~\msun\ is proposed by Valenti \& Fischer 2005), we note that this star is also characterized by a short rotation period, which might help to explain the shortness of its activity cycle (with a  length of 2 years only, Fares et al. 2009).

\begin{acknowledgements}
We thank the staff of TBL for their help during this observing run. V. Van Grootel acknowledges grant support from the Centre National d'Etudes Spatiales (CNES, France). We are grateful to the referee, Dr Stephen Marsden, for a number of comments that helped to improve this article.
\end{acknowledgements}


\begin{thebibliography}{}

 \bibitem[2008]{baliunas95} Baliunas SL, Donahue RA, Soon WH, Horne JH, Frazer J, Woodard-Eklund L, Bradford M, Rao LM, Wilson OC, Zhang Q, el al., 1995, ApJ 438, 269
\bibitem[2009]{brown09} Brown BP, Browning MK, Miesch MS, Brun AS, Toomre J. 2009, arXiv:0906.2407
 \bibitem[2008]{chandrasekhar61} Chandrasekhar S., 1961, Hydrodynamic and Hydromagnetic Stability. International Series of Monographs on Physics, Oxford, Clarendon
 \bibitem[1997]{donatietal97} Donati, J.F., Semel, M., Carter, B.D., Rees, D.E. \& Collier Cameron, A., 1997, MNRAS 291, 658
 \bibitem[1997]{donatibrown97} Donati, J.-F. \& Brown, S.F., 1997, A\&A 326, 1135
 \bibitem[2008]{donati06} Donati J-F, Howarth ID, Jardine MM, Petit P, et al., 2006, MNRAS, 370, 629
 \bibitem[2008]{fares09} Fares, R.; Donati, J. -F.; Moutou, C.; Bohlender, D.; Catala, C.; Deleuil, M.; Shkolnik, E.; Cameron, A. C.; Jardine, M. M.; Walker, G. A. H., 2009, MNRAS 398, 1383
 \bibitem[2008]{hall07a} Hall JC, Lockwood GW \& Skiff BA, 2007a, AJ 133, 862
 \bibitem[2008]{hall07b} Hall JC, Henry GW \& Lockwood GW, 2007b, AJ 133, 2206
 \bibitem[2007]{moutou2007}  Moutou, C., Donati, J.-F., Savalle, R., et al., 2007, A\&A, 473, 651
 \bibitem[2008]{nidever05} Nidever, D. L., Marcy, G. W., Butler, R. P., Fischer, D. A., \& Vogt, S. S., 2002, ApJS, 141, 503
 \bibitem[2002]{petit02} Petit, P., Donati, J.-F. \& Collier Cameron, A., 2002, MNRAS 334, 374
 \bibitem[2008]{petit05}  Petit P, Donati JF, Auri\`ere M, Landstreet JD, Ligni\`eres F, Marsden S, Mouillet D, Paletou F, Toqu\'e N, Wade GA, 2005, MNRAS 361, 837
 \bibitem[2008]{petit08} Petit, P., Dintrans, B., Solanki, S.K., et al., 2008, MNRAS 388, 80
 \bibitem[2008]{sanderson03} Sanderson TR, Appourchaux T, Hoeksema JT, Harvey KL, 2003, JGR, Vol. 108, Issue A1, (SSH 7-1)
 \bibitem[2008]{schrijver89} Schrijver CJ, Cote J, Zwaan C, Saar SH, 1989, AJ, 337, 964
 \bibitem[2008]{skilling84} Skilling J., Bryan R. K., 1984, MNRAS, 211, 111
 \bibitem[2005]{valenti05} Valenti, J.F., Fischer, D.A., 2005, ApJS 159, 141
\bibitem[2004]{wright04} Wright, J.T., Marcy, G.W., Butler, R.P. \& Vogt, S.S., 2004, ApJS, 152, 261

\end{thebibliography}
\end{document}